\documentclass{article}
\usepackage{times}
\usepackage{amsfonts}
\usepackage{graphicx}
\begin{document}
\noindent
{\Large  THE COSMOLOGICAL CONSTANT OF EMERGENT\\ SPACETIME IN THE NEWTONIAN APPROXIMATION}
\noindent

\vskip.5cm
\noindent
{\bf J.C. Castro-Palacio}$^{a}$, {\bf P. Fern\'andez de C\'ordoba}$^{b}$ and {\bf  J.M. Isidro}$^{c}$\\
Instituto Universitario de Matem\'atica Pura y Aplicada,\\ Universidad Polit\'ecnica de Valencia, Valencia 46022, Spain\\
${}^{a}${\tt juancas@upvnet.upv.es}, ${}^{b}${\tt pfernandez@mat.upv.es},\\
${}^{c}${\tt joissan@mat.upv.es}
\vskip.5cm
\noindent
\today
\vskip.5cm
\noindent
{\bf Abstract}  We present a simple quantum--mechanical estimate of the cosmological constant of a Newtonian Universe. We first mimic the dynamics of a Newtonian spacetime by means of a nonrelativistic quantum mechanics for the matter contents of the Universe (baryonic and dark)  within a fixed ({\it i.e.}\/, nondynamical) Euclidean spacetime. Then we identify an operator that plays, on the matter states, a role analogous to that played by the cosmological constant. Finally we prove that there exists a quantum state for the matter fields, in which the above mentioned operator has an expectation value equal to the cosmological constant of the given Newtonian Universe.

\section{Introduction}\label{einfuehrung}

\subsection{Goals}

While avoiding the sophistication of general relativity, Newtonian cosmology succeeds in capturing some essential physics of the Universe. In this letter we take advantage of this simplification in order to present a toy model of a Newtonian Universe, in which a certain repulsive force plays a role analogous to that played by the cosmological constant $\Lambda$ in general relativity. 

In natural units one has the approximate value $\Lambda\simeq 10^{-122}$ \cite{PLANCK}. Therefore  $\Lambda^{-1}\simeq 10^{122}$, a number that comes surprisingly close to $S_{\rm max}/k_B\simeq 10^{123}$. Here $k_B$ denotes the Boltzmann constant, while $S_{\rm max}$ is the maximal entropy of the Universe allowed by the holographic principle \cite{THOOFT, SUSSKIND}. This approximate equality of two dimensionless quantities might be a coincidence, but it might also hide some deeper connection. In this paper we explore this latter possibility. 

In refs. \cite{PADDY1, PADDY2}, a nonzero value of the cosmological constant has been argued to render the amount of cosmic information accessible to an eternal observer finite. This relation has allowed to determine the numerical value of the cosmological constant from the quantum structure of spacetime. Information is of course related to entropy, which seems to point towards the existence of some link between the two dimensionless numbers $\Lambda^{-1}\simeq 10^{122}$ and $S_{\rm max}/k_B\simeq 10^{123}$.

Of course one would like to be able to derive the above numbers from a full--fledged theory of quantum gravity.  More modestly, in refs.  \cite{CABRERA, NOI} we have obtained $S_{\rm max}/k_B\simeq 10^{123}$ using a finite--dimensional quantum--mechanical model\footnote{By this we mean the quantum theory of a {\it finite}\/ number of degrees of freedom, as opposed to a field theory.} of a Newtonian Universe. This strongly suggests that the cosmological constant $\Lambda$ must also be obtainable from the same model as in refs. \cite{CABRERA, NOI}. Carrying out this computation is one of the goals of the present paper. 

Another goal is to establish a link between the operator that represents the cosmological constant $\Lambda$, on the one hand, and the entropy operator $S$ used in refs. \cite{CABRERA, NOI} to compute the number $S_{\rm max}/k_B\simeq 10^{123}$, on the other. For this we will resort, as already done in refs. \cite{CABRERA, NOI}, to some entropic notions concerning the emergent nature of spacetime put forward in ref. \cite{VERLINDE}.

\subsection{The cosmological constant as an operator}

The Newtonian gravitational attraction created by a unit point mass located at the origin is described by the attractive potential $-1/r$.  The derivative of the latter, $1/r^2$, is of course (minus) the Newtonian force. But one can also interpret $1/r^2$ as yet another potential, the corresponding  {\it repulsive}\/ force decaying like $1/r^3$. In this letter we take the point of view that the repulsive potential $1/r^2$ plays the role of a cosmological constant within a hypothetical Newtonian Universe. Arguments in favour of this interpretation are given below.

{}Following the steps of refs. \cite{CABRERA, NOI}, we will model the volume density of (baryonic and dark) mass in this toy Universe as being given by $\vert\psi({\bf x}, t)\vert^2$. Here $\psi({\bf x}, t)$ is a quantum wavefunction satisfying a 1--particle Schroedinger equation. The value of the mass $m$ present in this equation is the total (baryonic and dark) matter contents of the Universe; the mass $m_V$ enclosed by a volume $V$ equals $m_V=m\int_V{\rm d}^3x\vert\psi\vert^2$. By the correspondence principle, this quantum--mechanical model exhibits classical properties in the limit of large quantum numbers.  We will {\it a posteriori}\/ verify that the relevant quantum numbers involved do indeed fall within the semiclassical regime.

In the literature, the cosmological constant $\Lambda$ has adopted a number of different impersonations. To name but a few, in ref. \cite{RUSOS} is has appeared as the eigenvalue of a Sturm--Liouville problem, while in ref. \cite{BARROW} it has been regarded as a field. In our finite--dimensional  quantum--mechanical model an operator $\Omega$ and a state $\vert\psi\rangle$ can be identified, such that the expectation value $\langle\psi\vert\Omega\vert\psi\rangle$ will equal $\Lambda$. Specifically, the sought--for operator $\Omega$ is 
\begin{equation}
\Omega=\frac{1}{r^2};
\label{landaerre}
\end{equation}
the state $\vert\psi\rangle$ will be specified presently. Up to a dimensionless multiplicative constant, the above form of $\Omega$ is dictated by symmetry (spatial isotropy) and by dimensional arguments: in standard ({\it i.e.}\/, nonnatural) units, $\Lambda$ has the dimensions of inverse length squared. We  see that $\Omega$ differs from the usual centrifugal potential $\hbar^2l(l+1)/(2mr^2)$ by multiplicative factors. However the centrifugal potential has a vanishing expectation value in the spherically symmetric $l=0$ states. This allows us to regard $\Omega$ as an analogue of the centrifugal force that however does not vanish in the $l=0$ states. 

Based on the operator (\ref{landaerre}), our estimates for the cosmological constant $\Lambda$ are given by the dimensionless numbers $\Lambda_n$ defined as
\begin{equation}
\Lambda_n:=\frac{\hbar}{mH_0}\langle\psi_n\vert\Omega\vert\psi_n\rangle, \qquad n=1,2,\ldots,
\label{naddall}
\end{equation}
the best fit to the experimentally measured value of $\Lambda$ being obtained for a certain value of the quantum number $n$ in the semiclassical regime. Here the $\{\psi_n\}_{n=1}^{\infty}$ form a complete, orthonormal set of states in Hilbert space, $H_0$ is Hubble's constant, $m$ is the (baryonic and dark) mass of the observable Universe, and $R_0$ its radius. The dimensionless combination $\lambda:=\hbar/(mH_0R_0^2)$ will appear frequently; its numerical value is \cite{PLANCK}
\begin{equation}
\lambda:=\frac{\hbar}{mH_0R_0^2}=2.6\times 10^{-124}.
\label{kicktrumpout}
\end{equation}
The dimensionful constants $\hbar$, $m$ and $H_0$ are imposed on us by the requirement that $\Lambda$ be dimensionless (when expressed in natural units). Three such constants are needed, with the dimensions of action, mass and frequency; respectively Planck's constant, the (baryonic and dark) matter contents of the Universe, and Hubble's constant are natural choices to make.

Altogether, in units of the dimensionless parameter $\lambda$, our value $\Lambda_n$ for the cosmological constant $\Lambda$ is
\begin{equation}
\Lambda_n=\lambda R_0^2\langle\psi_n\vert\Omega\vert\psi_n\rangle, \qquad n=1,2,\ldots
\label{yoko}
\end{equation}
We will compute $\Lambda_n$ with respect to the eigenstates $\vert\psi_n\rangle$ of the free, 1--particle Hamiltonian operator
\begin{equation}
H_{\rm F}=-\frac{\hbar^2}{2m}\nabla^2,
\label{frei}
\end{equation}
which will be assumed to describe the matter contents of our hypothetical Newtonian Universe. We will find that the best fit of $\Lambda_n$ to the measured value of $\Lambda$  \cite{PERLMUTTER, RIESS} will occur for a certain value of $n$ in the semiclassical regime. However, in the first place, we need to justify our use of finite--dimensional quantum mechanics in the study of a Newtonian cosmological problem.

\section{Newtonian cosmology as a quantum mechanics}

In Newtonian cosmology (for a summary see, {\it e.g.}\/, ref. \cite{BONDI}) the Universe is regarded as being (a submanifold of) Euclidean $\mathbb{R}^3$. The gravitational potential $U$ satisfies the Poisson equation 
\begin{equation}
\nabla^2U=4\pi G\rho.
\label{ztretre}
\end{equation}
The matter content (baryonic and dark) is modelled as an ideal fluid satisfying the continuity equation and the Euler equation,
\begin{equation}
\frac{\partial\rho}{\partial t}+\nabla\cdot\left(\rho{\bf v}\right)=0,\qquad \frac{\partial{\bf v}}{\partial t}+\left({\bf v}\cdot\nabla\right){\bf v}+\frac{1}{\rho}\nabla p-{\bf F}=0.
\label{zknoott}
\end{equation}
The cosmological principle requires that the velocity field ${\bf v}$ be everywhere proportional to the position vector ${\bf r}$. This requirement is equivalent to Hubble's law \cite{HUBBLE}, which can be described phenomenologically by the harmonic potential
\begin{equation}
U_{\rm Hubble}({\bf r})=-\frac{H_0^2}{2}{\bf r}^2. 
\label{zpotenzi}
\end{equation}
Hubble's constant $H_0$ is a frequency; the negative sign implies that this potential is repulsive. Accordingly, $U_{\rm Hubble}$ satisfies the Poisson equation (\ref{ztretre}) with a {\it negative}\/ mass density.

Although not widely recognised, Schroedinger quantum mechanics can also be understood in terms of an ideal fluid, the {\it quantum probability fluid}\/. Following Madelung \cite{MADELUNG} one factorises the nonrelativistic wavefunction $\psi$ into amplitude and phase:
\begin{equation}
\psi=\exp\left(\frac{{\cal S}}{2k_B}+{\rm i}\frac{{\cal I}}{\hbar}\right).
\label{zbenouere}
\end{equation}
The amplitude $\exp({\cal S}/2k_B)$ is a real exponential; one can invoke Boltzmann's principle to regard  ${\cal S}$ as a Boltzmann entropy of the matter described by $\psi$. It will also be convenient to define a dimensionless Boltzmann entropy $S:={\cal S}/2k_B$. The phase $\exp({\rm i}{\cal I}/\hbar)$ is the complex exponential of the classical--mechanical action integral ${\cal I}$. Substituting the Ansatz (\ref{zbenouere}) into the Schroedinger equation for $\psi$, one arrives at a set of two equations. One of them is the continuity equation for the quantum probability fluid,
\begin{equation}
\frac{\partial S}{\partial t}+\frac{1}{m}\nabla S\cdot\nabla {\cal I}+\frac{1}{2m}\nabla^2{\cal I}=0,
\label{konntt}
\end{equation}
where 
\begin{equation}
{\bf v}:=\frac{1}{m}\nabla {\cal I}, \qquad \rho={\rm e}^{2S}.
\label{zvanednie}
\end{equation}
The second equation obtained is known as the {\it quantum Hamilton--Jacobi equation}\/:
\begin{equation}
\frac{\partial {\cal I}}{\partial t}+\frac{1}{2m}(\nabla {\cal I})^2+V+{\cal Q}=0,
\label{zvabueelnv}
\end{equation}
where $V$ is the external potential present in the Schroedinger equation.\footnote{We recall that the dimensions of $U$ in (\ref{ztretre}) and (\ref{zpotenzi}) are velocity squared, whereas those of $V$ in (\ref{zvabueelnv}) are mass times velocity squared.} Above,  
\begin{equation}
{\cal Q}:=-\frac{\hbar^2}{2m}\left[\left(\nabla S\right)^2+\nabla^2 S\right]
\label{zvebiyer}
\end{equation}
is known as the {\it quantum potential}\/ \cite{MATONE}. Indeed, in the limit when $\hbar\to 0$, the quantum potential vanishes, and (\ref{zvabueelnv}) reduces to the classical Hamilton--Jacobi equation.

In order to be able to interpret quantum mechanics as an Euler fluid we need to derive an Euler equation for the quantum probability fluid. This is achieved by taking the gradient of  Eq. (\ref{zvabueelnv}):
\begin{equation}
\frac{\partial{\bf v}}{\partial t}+\left({\bf v}\cdot\nabla\right){\bf v}+\frac{1}{m}\nabla {\cal Q}+\frac{1}{m}\nabla V=0.
\label{kblm}
\end{equation}
Comparison between Eqs. (\ref{kblm}) and (\ref{zknoott}) produces a bijective correspondence between the quantum probability fluid and the cosmological fluid. According to this correspondence, Newtonian cosmology can be regarded in either one of two equivalent descriptions:
\begin{equation}
\begin{tabular}{| c | c | c|}\hline
& Euler &  Madelung  \\ \hline
fluid density & $\rho$ & $\exp(2S)$   \\ \hline
velocity field & ${\bf v}$ & $\nabla{\cal I}/m$   \\ \hline
fluid pressure & $\nabla p/\rho$ & $\nabla{\cal Q}/m$   \\ \hline
force per unit mass & ${\bf F}$ & $-\nabla V/m$   \\ \hline
\end{tabular}
\label{tabulauno}
\end{equation}
Which suggests that, {\it given the cosmological fluid in the Newtonian approximation, we use nonrelativistic quantum mechanics as an equivalent description thereof}\/. 

We would like to add that, beyond Newtonian cosmology, the dark Universe has also found an interesting description as a cosmological fluid in refs. \cite{BOLOGNA1, BOLOGNA2, BOLOGNA3}.

\section{Computing $\Lambda$}\label{trumpburro}

The free Hamiltonian (\ref{frei}) admits  the spherical waves
\begin{equation}
\psi_{\kappa 00}(r,\theta,\phi)=\frac{1}{\sqrt{4\pi R_0}}\frac{1}{r}\exp\left({\rm i}\kappa r\right), \qquad\kappa\in\mathbb{R}
\label{iggn}
\end{equation}
as eigenfunctions. They are normalised within a sphere of radius $R_0$ (the radius of the observable Universe); $\vert\kappa\vert$ is the modulus of the linear momentum; the angular quantum numbers $l,m$ have been set to zero as demanded by the cosmological principle. For regularity of the wavefunction at the origin $r=0$ we will consider the normalised linear combination of the states (\ref{iggn}) given by
\begin{equation}
\psi_n(r)=\frac{1}{\sqrt{2\pi R_0}}\frac{1}{r}\sin\left(\frac{n\pi}{R_0} r\right), \qquad n=1,2,\ldots
\label{psiene}
\end{equation}
where the boundary condition $\psi(R_0)=0$ has been imposed. One readily finds the expectation value of the operator (\ref{landaerre}):
\begin{equation}
\langle\psi_n\vert\Omega\vert\psi_n\rangle=\frac{2\pi n}{R_0^2}{\rm Si}\left(2n\pi\right),\qquad n=1,2,\ldots
\label{sekko}
\end{equation}
where ${\rm Si}(x)=\int_0^x{\rm d}t\sin(t)/t$ \cite{LEBEDEV}. By Eq. (\ref{sekko}) in (\ref{yoko}) we arrive at the estimates $\Lambda_n$ for the cosmological constant $\Lambda$:
\begin{equation}
\Lambda_n=\lambda f(n),\qquad f(n):=2n\pi\,{\rm Si}\left(2n\pi\right), \qquad n=1,2,\ldots
\label{stoptrump}
\end{equation}
Evaluating $f(n)$ we can tabulate the first few values of $\Lambda_n$:
\begin{equation}
\begin{tabular}{| c | c | c|}\hline
$\Lambda_n$ & in units of $\lambda$ &  in natural units  \\ \hline
$n=1$ & $8.9$ & $2.3\times 10^{-123}$   \\ \hline
$n=2$ & $18.8$ & $4.9\times 10^{-123}$   \\ \hline
$n=3$ & $28.6$ & $7.4\times 10^{-123}$ \\ \hline
$n=4$ & $38.5$ & $1.0\times 10^{-122}$ \\ \hline
$n=5$ & $48.4$ & $1.3\times 10^{-122}$ \\ \hline
$n=6$ & $58.2$ & $1.5\times10^{-122}$ \\ \hline
$n=7$ & $68.1$ & $1.8\times 10^{-122}$ \\ \hline
$n=8$ & $78.0$ & $2.0\times 10^{-122}$ \\ \hline
$n=9$ & $87.8$ & $2.3\times 10^{-122}$ \\ \hline
$n=10$ & $97.7$ & $2.5\times 10^{-122}$ \\ \hline
$n=11$ & $107.6$ & $2.8\times 10^{-122}$ \\ \hline
$n=12$ & $117.4$ & $3.0\times 10^{-122}$ \\ \hline
\end{tabular}
\label{tabulados}
\end{equation}
They are in excellent agreement with the measured value of $\Lambda$, the best fit being around $n=11$, well inside the semiclassical regime as announced. As we let $n\to\infty$ we find
\begin{equation}
\langle\psi_n\vert\Omega\vert\psi_n\rangle\simeq\frac{n\pi^2}{R_0^2},\qquad\Lambda_n\simeq\lambda\pi^2n.
\label{gadde}
\end{equation}

\section{The cosmological constant operator as the inverse to the entropy operator}

\subsection{Emergent spacetime}

In  a Newtonian Universe, the area measured by $r^2$ has been argued to be proportional to the entropy within the volume enclosed by the given area. This point has been discussed at length in ref. \cite{VERLINDE}, where an entropic concept for the emergent nature of spacetime has been put forward.  

The operator $r^2$ has played a key role in our computation of the number $S_{\rm max}/k_B\simeq 10^{123}$ in refs. \cite{CABRERA, NOI}. Specifically, the holographic bound $S_{\rm max}/k_B\simeq 10^{123}$ has been achieved as the expectation value of the operator $S=k_BmH_0 r^2/\hbar$ in a certain quantum eigenstate of the Hamiltonian (\ref{frei}). Modulo the physical constants $k_B$ $m$, $H_0$ and $\hbar$, the entropy operator used in refs. \cite{CABRERA, NOI} is the inverse operator to $1/r^2$ used here. 

Therefore entropy and the cosmological constant are mutually inverse, not just as numbers as mentioned in section \ref{einfuehrung}, but also as quantum--mechanical operators.  The explanation for the approximate equality between the dimensionless numbers $\Lambda^{-1}\simeq 10^{122}$ and $S_{\rm max}/k_B\simeq 10^{123}$ lies in the deeper fact that {\it the corresponding quantum--mechanical operators are mutually inverse}\/. 

Now the operator $\Omega=1/r^{2}$ becomes singular at $r=0$. Below we explore the consequences of this singularity in more detail. As we will see, one surprising consequence of the regularisation of this singularity will be the natural appearance of a quantum of length.

\subsection{Regularisation}\label{hadd}

Consider the coefficients $c_n$ of the Fourier--Bessel expansion of $\Omega=1/r^2$ in terms of Bessel functions \cite{LEBEDEV}:
\begin{equation}
c_n=\frac{2}{[R_0J'_{p}(z_{p}(n))]^2}\int_0^{R_0}J_{p}\left(\frac{z_{p}(n)}{R_0}r\right)\frac{{\rm d}r}{r},\qquad n=1,2,\ldots
\label{samschweber}
\end{equation}
Here $J'_p(x)$ denotes the derivative of the Bessel function $J_p(x)$ and $z_{p}(n)$ denotes the positive zeroes of the latter; the value of $p$ is specific to the precise Hamiltonian considered ($p=1/2$ for the Hamiltonian (\ref{frei})).  As will be proved below, it turns out that our $\Lambda_n$ of Eq. (\ref{yoko}) are proportional to the Fourier--Bessel coefficients $c_n$ in the limit when $n\to\infty$.

{}From a knowledge of the $c_n$ one may construct the formal series
\begin{equation}
\sum_{n=1}^{\infty}c_nJ_{p}\left(\frac{z_{p}(n)}{R_0}r\right).
\label{pentwicklung}
\end{equation}
Unfortunately the function $1/r^2$ on $(0,R_0)$ does not satisfy the conditions required for the series (\ref{pentwicklung}) to converge to the function $1/r^2$ on $(0,R_0)$ \cite{LEBEDEV}. However this does not deprive the coefficients $c_n$ of their meaning, since Eq. (\ref{samschweber}) make sense {\it per se}\/. Moreover, the cosmological constant  being relevant only at astronomical scales, $\Lambda$ does not affect the short--range physics. One could then introduce an ultraviolet cutoff $\epsilon>0$ and consider the function $1/r^2$ on $(\epsilon, R_0)$, where it does satisfy the requisite conditions for the series (\ref{pentwicklung}) to converge to $1/r^2$. The integrals in (\ref{samschweber}) would now extend over the interval $(\epsilon, R_0)$. Whichever interpretation one chooses (either to ignore convergence issues or to enforce convergence through an ultraviolet cutoff), the fact remains that, {\it in the semiclassical limit, the Fourier--Bessel coefficients of $1/r^2$ are proportional to the expectation values  of $1/r^2$}\/, as we establish next.

The complete orthonormal system of radial states (\ref{psiene}) can be recast in terms of Bessel functions as
\begin{equation}
\psi_n(r)=\frac{\sqrt{n\pi}}{2R_0}\frac{1}{\sqrt{r}}J_{\frac{1}{2}}\left(\frac{n\pi}{R_0}r\right),\quad n=1,2,\ldots
\label{eijenn}
\end{equation}
Thus setting $p=1/2$ and $z_{1/2}(n)=n\pi$, the Fourier--Bessel coefficients  (\ref{samschweber}) now read
\begin{equation}
c_n=\frac{2}{[R_0J'_{1/2}(n\pi)]^2}\int_0^{R_0}J_{\frac{1}{2}}\left(\frac{n\pi}{R_0}r\right)\frac{{\rm d}r}{r}=\frac{4n\pi^2}{R_0^2}C\left(\sqrt{2n}\right),
\label{schweber}
\end{equation}
where $C(z)=\int_0^z\cos\left(\pi t^2/2\right){\rm d}t$ \cite{LEBEDEV}. In the limit $n\to\infty$ we have
\begin{equation}
c_n\simeq \frac{2n\pi^2}{R_0^2}.
\label{adieu}
\end{equation}
Comparing Eqs. (\ref{adieu}) and (\ref{gadde}) we observe that, in the semiclassical limit, the Fourier--Bessel coefficients  of $1/r^2$ are indeed proportional to $\langle\psi_n\vert\Omega\vert\psi_n\rangle$ as announced.

{}Finally we repeat the above calculations without imposing a boundary condition at $r=R_0$. In particular this implies that the wavefunctions (\ref{iggn}) are permitted, even though they are singular at $r=0$. We have
\begin{equation}
\langle\psi_{\kappa}\vert \Omega\vert\psi_{\kappa}\rangle=\frac{1}{R_0}\int_0^{R_0}\frac{{\rm d}r}{r^2}
\label{avignon}
\end{equation}
This integral diverges at its lower limit; this is a consequence of the singular behaviour, at $r=0$, of both the wavefunction $\psi_{\kappa}$ and the operator $\Omega$. We regularise the integral (\ref{avignon}) by simply dropping the infinite contribution coming from its lower limit and find a regularised expectation value
\begin{equation}
\langle\psi_{\kappa}\vert \Omega\vert\psi_{\kappa}\rangle_{\rm reg}=-\frac{1}{R_0^2}.
\label{reku}
\end{equation}
While the operator $\Omega$ is nonnegative, a surprising minus sign appears on the right--hand side; this is a consequence of the regularisation.\footnote{This is not unsual. A well--known example is the $\zeta$--function regularisation of the divergent sum of positive integers $\sum_{n=1}^{\infty}n$, which yields the {\it negative}\/ value $-1/12$ \cite{ELIZALDE}.} With the regularised integral (\ref{reku}) we return to Eq. (\ref{naddall}) and, changing the sign,  find the dimensionless value
\begin{equation}
\langle\Lambda\rangle_{\rm reg}=-\frac{\hbar}{mH_0}\langle\Omega\rangle_{\rm reg}=\frac{\hbar}{mH_0R_0^2},
\label{rkki}
\end{equation}
in perfect agreement with our previous result (\ref{kicktrumpout}).

The above regularisation procedure may appear somewaht {\it ad hoc}\/, but in fact it is not. It is known in the literature under the dignified name of  Hadamard regularisation \cite{BLANCHET}, a popular choice when regularising  gravitational models. In what follows we briefly summarise the Hadamard regularisation in order to provide an alternative derivation of our Eq. (\ref{reku}). Given $x\in(a,b)$ and a function $f(t)$ sufficiently differentiable at $t=x$, consider the divergent integral
\begin{equation}
\int_a^b\frac{f(t)}{(t-x)^2}{\rm d}t.
\label{cochy}
\end{equation}
Then the Hadamard {\it partie finie}\/ of (\ref{cochy}), denoted by an ${\cal H}$ in front,  is defined by
\begin{equation}
{\cal H}\int_a^b\frac{f(t)}{(t-x)^2}{\rm d}t
\label{hadamard}
\end{equation}
$$
=\lim_{\epsilon\to 0^+}\left[\int_a^{x-\epsilon}\frac{f(t)}{(t-x)^2}{\rm d}t+\int_{x+\epsilon}^b\frac{f(t)}{(t-x)^2}{\rm d}t-\frac{f(x-\epsilon)+f(x+\epsilon)}{\epsilon}\right]
$$
In the case at hand we rewrite our divergent integral (\ref{avignon}) as 
\begin{equation}
\langle\psi_{\kappa}\vert \Omega\vert\psi_{\kappa}\rangle=\frac{1}{2R_0}\int_{-R_0}^{R_0}\frac{{\rm d}r}{r^2}
\label{surlepontdavignon}
\end{equation}
in order to identify $a=-R_0$, $b=R_0$, $x=0$, $f(t)=1$. Applying Eq. (\ref{hadamard}) one readily verifies that
\begin{equation}
{\cal H}\int_{-R_0}^{R_0}\frac{{\rm d}r}{r^2}=-\frac{2}{R_0}.
\label{reggu}
\end{equation}
{}Finally Eqs. (\ref{surlepontdavignon}) and (\ref{reggu}) yield back our former result (\ref{reku}). We conclude that the ultraviolet cutoff $\epsilon>0$ mentioned previously is nothing but Hadamard regularisation.

\section{Discussion}

We have presented a nonrelativistic, quantum--mechanical model of a hypothetical Newtonian Universe, in which an operator can be identified that plays a role analogous to a cosmological constant. Indeed the expectation value of this operator in a certain basis of quantum states yields a value that lies close to the measured value of the cosmological constant.

Admittedly, our toy model is not a theory of quantum spacetime; it contains one adjustable parameter, the quantum number $n$. All this notwithstanding, the close equality of the two dimensionless quantities $\Lambda^{-1}\simeq 10^{122}$ and $S_{\rm max}/k_B\simeq 10^{123}$ has a very simple explanation in our framework. Namely, the value $\Lambda^{-1}\simeq 10^{122}$ arises as the (inverse of the) expectation value of the operator $1/r^2$, while the $S_{\rm max}/k_B\simeq 10^{123}$ arises as the expectation value of $r^2$. Both expectation values are taken within the {\it same}\/ quantum--mechanical model. Moreover, the Hadamard regularisation used in section \ref{hadd}, seen to be equivalent to the inclusion of a cutoff $\epsilon$ in coordinate space, can be regarded as an indication of the existence of a quantum of length $\epsilon$. Thus the requirement of {\it finiteness}\/ of the cosmological constant leads naturally to the existence of a quantum of length.

Our result for $\Lambda$ depends on the specific choice of a complete set of orthonormal vectors $\left\{\vert\psi_n\rangle\right\}_{n=1}^{\infty}$ for 1--particle states; it also depends on the quantum number $n$. We have established that, for a given choice of orthonormal vectors, the best fit of $\Lambda_n$ to the measured value of $\Lambda$ occurs for a certain value of $n$ in the semiclassical regime. Of course, different Hamiltonians will lead to different choices for the set $\left\{\vert\psi_n\rangle\right\}_{n=1}^{\infty}$, and correspondingly to different values of $\Lambda_n$. In previous work by some of the present authors \cite{CABRERA, NOI, FINSTER}  we have modelled the Hubble expansion of a Newtonian Universe by means of an inverse oscillator \cite{PADDY3}
\begin{equation}
H_{\rm H}=H_{\rm F}-\frac{k_{\rm eff}r^2}{2},\qquad k_{\rm eff}=mH_0^2.
\label{jabel}
\end{equation}
In upcoming work we propose to use the Hamiltonian (\ref{jabel}) in order to refine our present estimate for the cosmological constant of a Newtonian Universe.

\vskip0.5cm
\noindent
{\bf Acknowledgements} This research was supported by grant no. RTI2018-102256-B-I00 (Spain).

\end{document}